\documentclass[usenatbib]{mnras}
\usepackage[dvips]{color}
\usepackage{ulem}
\usepackage{longtable}
\usepackage{url}

\usepackage{graphicx}
\usepackage{lscape}

\def\ergscm{erg~s$^{-1}$~cm$^{-2}$}

\def\arcmin{\hbox{$^\prime$}}
\def\arcsec{\hbox{$^{\prime\prime}$}}

\def\flux{erg s$^{-1}$ cm$^{-2}$}
\def\lum{erg s$^{-1}$}

\def\apj{ApJ}
\def\apss{Astroph.Sp.Sci.}
\def\aap{A\&A}
\def\mnras{MNRAS}
\def\apjs{ApJS}

\def\apjl{ApJLett}

\def\gro{GRO\,J1750$-$27}
\def\2s{2S\,1553-542}

\newcommand {\be}{\begin {equation}}
\newcommand {\ee}{\end {equation}}

\title[GROJ1750-27: neutron star behind Galactic Center]
{GRO\,J1750$-$27: a neutron star far behind the Galactic Center switching into the propeller regime}
\author[Lutovinov et al.]{Alexander A.\,Lutovinov$^{1}$\thanks{E-mail: aal@iki.rssi.ru},
Sergey S.\,Tsygankov$^{2,1}$, Dmitri I.\,Karasev$^{1}$,
\newauthor Sergei V.\,Molkov$^{1}$ and Viktor Doroshenko$^{3}$
\\
$^1$ Space Research Institute, Profsoyuznaya str. 84/32, Moscow, 117997, Russia  \\
$^2$ Department of Physics and Astronomy, University of Turku, FI-20014, Finland\\
$^{3}$ Institut f\"ur Astronomie und Astrophysik, Universit\"at T\"ubingen, Sand 1, D-72076 T\"ubingen, Germany}

\begin{document}

\date{Accepted .... Received ...}

\pagerange{\pageref{firstpage}--\pageref{lastpage}} \pubyear{2018}

\maketitle

\label{firstpage}

\begin{abstract}


{We report on analysis of properties of the X-ray binary pulsar \gro\ based
on X-ray ({\it Chandra}, {\it Swift}, and {\it Fermi}/GBM), and near-infrared
({\it VVV} and {\it UKIDSS} surveys) observations. An accurate position of
the source is determined for the first time and used to identify its infrared
counterpart. Based on the {\it VVV} data we investigate the spectral energy
distribution (SED) of the companion, taking into account a non-standard
absorption law in the source direction. A comparison of this SED with those
of known Be/X-ray binaries and early type stars has allowed us to estimate a
lower distance limit to the source at $>12$ kpc. An analysis of the observed
spin-up torque during a giant outburst in 2015 provides an independent
distance estimate of $14-22$\,kpc, and also allows to estimate the
magnetic field on the surface of the neutron star at $B\simeq(3.5-4.5)\times10^{12}$
G. The latter value is in agreement with the possible transition to the
propeller regime, a strong hint for which was revealed by {\it Swift}/XRT and
{\it Chandra}. We conclude, that \gro\ is located far behind the Galactic
Center, which makes it one of the furthest Galactic X-ray binaries known.}

\end{abstract}

\begin{keywords}
stars: individual: \gro -- X-rays: binaries
\end{keywords}

\section{Introduction}
\label{sec:intro}

High mass X-ray binaries (HMXBs), are binary systems hosting a compact object
(i.e. a neutron star, or a black hole), and massive non-degenerate companion.
A population of HMXBs is not uniform and can be roughly divided in several
types depending on the observable properties, evolutionary status and optical
companion \citep[see][for the recent review of HMXBs]{2015A&ARv..23....2W}.

A large fraction of HMXBs constitutes so-called Be/X-ray binaries (BeXRBs).
In these systems the optical companion is a fast rotating Be star with a
typical luminosity class III-V, demonstrating spectral emission lines, which
are formed in a rotating decretion disk around the star
\citep{2011Ap&SS.332....1R}. Most of BeXRBs have a transient nature in
X-rays, and demonstrate a strong variability of their X-ray luminosity of
several orders of magnitude from $L_{X} \sim 10^{33}$ \lum\ in quiescence, and up
to $L_{X} \sim 10^{39}$ \lum\ during the outbursts \citep[see,
e.g.,][]{2017A&A...605A..39T}. This makes such systems unique natural
laboratories for accretion studies \citep[see, e.g.,][and references
therein]{2001A&A...369..108N, 2001A&A...377..161O, 2013ApJ...777..115P,
2013PASJ...65...41O, 2017A&A...608A..17T}. A high X-ray luminosity of these
objects reaching during outbursts makes them also an important probe to study
the far side of the Galaxy \citep[see, e.g.,][]{2016MNRAS.462.3823L}.

The transient X-ray pulsar \gro\ was discovered with the BATSE instrument
onboard the {\it Compton-GRO} observatory during a strong outburst in July
1995 \citep{1995IAUC.6222....1K}. The source was detected and recognized as a
new X-ray pulsar due to a registration of coherent pulsations with the period of about
4.45 s, which was interpreted as a rotation period of the neutron star
\citep{1997ApJ...488..831S}. The outburst lasted about two months, which
allowed to determine not only orbital parameters of the system, but also to
investigate the spin-up evolution during the outburst.

A second outburst from \gro\ was observed in the beginning of 2008 with the
{\it Swift} and {\it INTEGRAL} observatories during several months
\citep{2009MNRAS.393..419S}. These observations allowed to improve the
orbital parameters of the system, to reconstruct for the first time a
broadband spectrum of the source in the 5-70 keV energy band and to obtain
estimates for the distance to \gro\ in the range of 12-22 kpc (based on the
spin-up evolution during the outburst). The corresponding peak luminosity was
estimated at the level of $\sim10^{38}$\,\lum. The high luminosity
and observed transient behaviour led to the conclusion that \gro\ should be a
Be/X-ray binary system demonstrating giant outbursts
\citep{1997ApJ...488..831S,2009MNRAS.393..419S}.

The sky region around \gro\ was observed in October 1995 with the {\it ASCA}
observatory, which detected the pulsations with the same period $\sim4.45$ sec
and improved the source localisation up to 2\arcmin\
\citep{1995IAUC.6241....2D}. However, the optical counterpart still could not
be unambiguously identified because the source is located in a crowded
Galactic Center field.

In this paper we report results of a comprehensive analysis of {\it Chandra,
Neil Gehrels Swift} and {\it Fermi} observations of \gro\ performed during
the latest outburst, started in December 2014. We obtained an accurate
localization of the source in X-rays and determined its infrared counterpart
for the first time, based on the {\it VVV} survey data. Using available
photometry of the counterpart, we estimated the spectral class of the optical
star and distance to the system. We also traced evolution of the source X-ray
flux and spectral parameters throughout the outburst, and discovered a strong
hint for the transition of the source into a propeller regime.

\section{Observations and data analysis}
\label{sec:obs}

\gro\ was monitored in soft X-rays with the {\it Swift}/XRT telescope during
both outbursts in 2008 and 2014-2015 (Target id. 31115). An important
difference between these monitoring programs is that during the 2008 outburst
the source was observed about dozen times in a high state, whereas observations
in 2015 were mostly done during the outburst decay from Apr 12 to May 20, 2015.

{\it Swift}/XRT observations were performed in the Windowed Timing (WT) and
Photon Counting (PC) modes depending on the source brightness. Final products
(spectrum in each observation) were prepared using online tools provided by
the UK Swift Science Data Centre
\citep{2009MNRAS.397.1177E}\footnote{\url{http://www.swift.ac.uk/user\_objects/}}.
Due to calibration uncertainties at low
energies\footnote{\url{http://www.swift.ac.uk/analysis/xrt/digest\_cal.php}},
we restricted the spectral analysis to the 0.7--10 keV energy band.

\begin{figure}
\centering
\includegraphics[width=0.95\columnwidth,bb=40 190 565 680,clip]{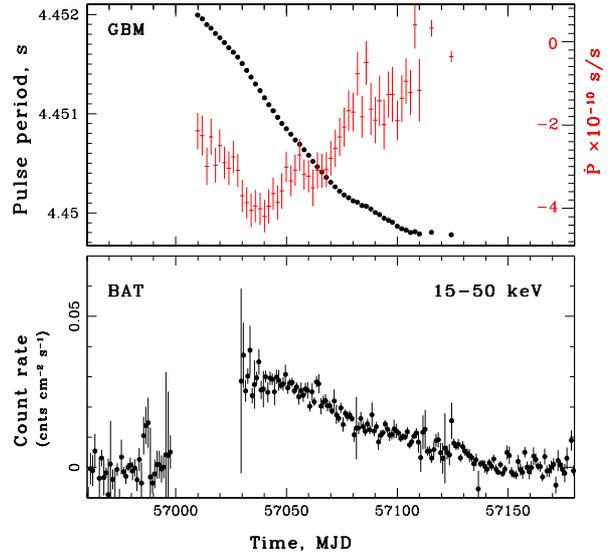}
\caption{Evolution of the pulse period and rate of the pulse period change
(upper panel) and hard X-ray count rate in the 15-50 keV energy band (bottom
panel) of \gro\ during the 2015 outburst.}\label{lcurve}
\end{figure}

The source was also monitored with the {\it Swift}/BAT
(\citealt{2013ApJS..209...14K}, 15-50 keV energy band) and {\it Fermi}/GBM
(\citealt{2009ApJ...702..791M}, 12-25 keV energy band) instruments. We used
these data to study the source behaviour at harder energies and to trace an
evolution of the pulse period and its rate of change during the outburst
(Fig.\ref{lcurve}). Data of both instruments were taken from the official
sites\footnote{\url{https://swift.gsfc.nasa.gov/results/transients/}},\footnote{\url{https://gammaray.nsstc.nasa.gov/gbm/science/pulsars/}}.

At the very end of the outburst (on May 20, 2015) {\it Chandra} observed
\gro\ with the ACIS instrument (ObsID. 16723) with a total exposure of
about 30 ks. The data were reduced with the standard software package
{\sc ciao 4.9}\footnote{\url{http://cxc.harvard.edu/ciao/}} with CALDB v4.7.5.

Taking into account the low flux of the source and low counting
statistics during {\it Swift}/XRT and {\it Chandra} observations the
obtained spectra were grouped to have at least 1 count per bin and were
fitted in the {\sc xspec} package using the $W$-statistic
\citep{1979ApJ...230..274W}.

It is important to note that for any meaningful discussions or
conclusions on the physical properties of the source a bolometric correction
has to be applied to the observed fluxes in soft or hard energy bands. After 
applying the bolometric correction, we recalculated the fluxes in XRT and BAT
energy bands to the energy range $0.1-100$ keV, which can be treated as a
bolometric one for X-ray pulsars with a good accuracy. Using the bolometric
flux of $6.5\times10^{-9}$\,\flux, spectral parameters and corresponding BAT
count rate from \citet{2009MNRAS.393..419S}, we estimate a conversion factor
between the observed BAT count rate and the bolometric flux as $K_{\rm
BAT}\simeq1.6\times10^{-7}$ \flux/(cnts s$^{-1}$). A conversion factor
between the flux, measured by XRT, and the bolometric flux was estimated as
$K_{\rm XRT}\simeq2.2$ from spectral shape and parameters, determined by
\citet{2009MNRAS.393..419S}. In the following analysis we apply these
correction to all observational data and refer to the bolometrically
corrected fluxes and luminosities, unless stated otherwise.

Finally, to search for the counterpart and to study its properties in the
infrared band we used the latest public releases of
{\it VVV}/ESO\footnote{\url{http://www.eso.org/sci/observing/phase3/data\_releases.html}},\footnote{\url{http://horus.roe.ac.uk/vsa/}} and
{\it GPS}/UKIDSS\footnote{\url{http://surveys.roe.ac.uk/wsa/}} sky survey data.

\section{Magnetic field and distance estimates from the spin-up measurements}

\label{sec:pdot}

\begin{figure}
\centering
\includegraphics[width=0.9\columnwidth,bb=40 175 550 680,clip] {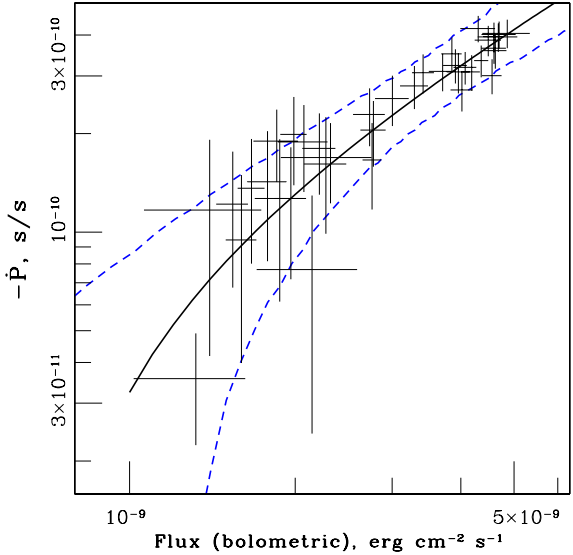}
\caption{Spin-up rate measured by {\it Fermi}/GBM depending on the bolometric
flux for \gro. Solid line shows the best approximation with the
Eq.(\ref{eq_pdot}). The area of possible solutions in frames of different
torque models \citep{1979ApJ...234..296G,1995ApJ...449L.153W,2007ApJ...671.1990K}
is restricted by dashed lines.} 
\label{pdot_flux}
\end{figure}

As it was mentioned above during the outburst we observe not only an
evolution of the \gro\ luminosity but also an evolution of the neutron star
spin-up rate. There are several models connecting a bolometric luminosity
$L_b$ of the source with its spinning-up rate $\dot P$ through the parameters
of the neutron star -- radius, mass, magnetic field, etc \citep[see,
e.g.][]{1979ApJ...234..296G,1995ApJ...449L.153W,2007ApJ...671.1990K,2016ApJ...822...33P}.
These models describe an interaction of the accretion disc with the
magnetosphere slightly differently, but for the bright outbursts they are
agreed well \citep[see, e.g.,][]{2016A&A...593A..16T,2017AstL...43..706F}.
Therefore in the following we assume that the $\dot P(L_b)$ dependence for
\gro\ is given by the equation (15) from the paper of
\citet{1979ApJ...234..296G}. We rewrite them in the form, which is more
convenient to describe observational data, using the bolometric flux ($F_b$)
instead the bolometric luminosity:

\be\label{eq_pdot}
-\dot P_{-10} \simeq 2.24~10^{-3} P^2 \mu_{30}^{2/7} D^{12/7} n(\omega_s) F_{b,-8}^{6/7}
\ee

\noindent where $ n(\omega_s)$ is a dimensionless accretion torque, describing the
``accretion disc -- magnetosphere'' interaction, $\dot P_{-10}$ is the
spin-up rate in units of $10^{-10}$ s\,s$^{-1}$, $P$ is the pulse period in seconds,
$\mu_{30}$ is the magnetic dipole moment in units of $10^{30}$~G~cm$^3$, $D$
is the distance to the source in kpc and $F_{b,-8}$ is the measured
bolometric flux in units of $10^{-8}$ \ergscm. We are assuming also that the
mass, radius and momentum of the inertia of the neutron star are
$1.4M_{\sun}$, $10^{6}$ cm and $10^{45}$~g\,cm$^2$, respectively.

The $\dot P(L_b)$ dependence for \gro, based on the {\it Swift}/BAT and {\it
Fermi}/GBM data, is plotted in Fig.\,\ref{pdot_flux}. Experimental data were
fitted with the Eq.(\ref{eq_pdot}) with best fit values of two parameters:
$D\simeq18$ kpc and $\mu_{30}\simeq2$, what corresponds to the strength of
the magnetic field on the neutron star surface $B\simeq4\times10^{12}$~G 
(we used here a relation of $\mu=BR^{3}/2$). This best fit model is shown by
a solid line in Fig.\,\ref{pdot_flux}. It is important to note, that this
estimate depends strongly on the assumed torque model, so the uncertainties
on these parameters can be quite large \citep[see Fig.\ref{pdot_flux} and,
e.g.,][]{2016A&A...593A..16T} and lead to the possible distance and magnetic
field ranges of $D\simeq14-22$ kpc and $B\simeq(3.5-4.5)\times10^{12}$~G.

\section{Propeller effect}
\label{sec:prop}

In addition to spin properties, the magnetic field of the accreting neutron
star can be estimated using large-scale variations of the observed flux. In
particular, detection the so-called ``propeller effect''
\citep{1975A&A....39..185I} can be used to estimate the field. The effect is
observed in highly magnetized compact objects due to an existence of the
critical mass accretion rate at which the magnetospheric radius $R_{\rm m}$
(which depends on the accretion rate), becomes comparable with the corotation
one $R_{\rm cor}$ (the radius where the linear velocity of the compact
object's rotation equals to the Keplerian one). At this point the velocity of
the magnetic field lines exceeds the velocity of the matter in the accretion
disc and the accretion is halted for low accretion rates due to an emerging
centrifugal barrier. From the observational point of view this effect is well
known for different types of binary systems with neutron stars ranging from
accreting millisecond and X-ray pulsars to accreting magnetars
\citep[][]{1986ApJ...308..669S,1997ApJ...482L.163C,2001ApJ...561..924C,
2008ApJ...684L..99C,
2016MNRAS.457.1101T,2016A&A...593A..16T,2017ApJ...834..209L,2018A&A...610A..46C}.
Recently, \citet{2017A&A...608A..17T} showed that only fast rotating neutron
stars (with the spin period of $P \la 10$~s) are able to enter the propeller
regime.

\begin{figure}
\includegraphics[width=0.95\columnwidth,bb=10 265 565 680,clip]{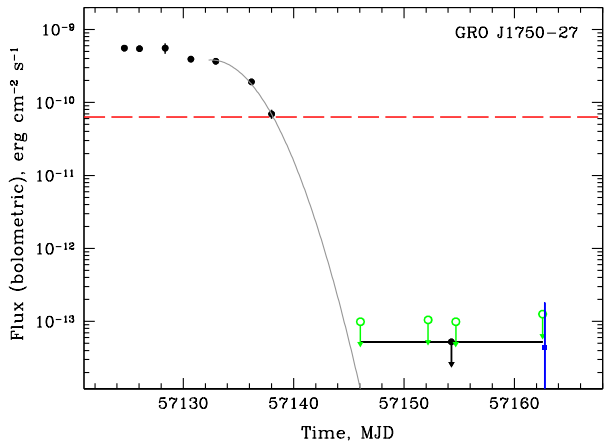}
\caption{Light curve of \gro\ measured with {\it Swift}/XRT and {\it
Chandra}. Black circles represent the bolometric unabsorbed flux measured
with the {\it Swift}/XRT telescope under assumption of the bolometric
correction factor of 2.2. Green open circles denote upper flux limits
for each of four {\it Swift}/XRT observations after the transition; an upper
limit corresponding to the total exposure of these observations is shown by
black circle with the arrow. The solid grey line illustrates the best fit of
the flux decay in the tail of the outburst with a Gaussian function.
Horizontal dashed line corresponds to the upper limit on the threshold flux
of the possible transition to the propeller regime.}\label{fig:lcxrt}
\end{figure}

Short spin period of \gro\ implies, that it must enter the propeller regime,
so its magnetic field can be estimated using a ``standard'' equation for the
limiting luminosity \citep[see,
e.g.,][]{2001ApJ...561..924C}:

\be\label{eq_prop} L_{\rm prop}
\simeq 4 \times 10^{37} k^{7/2}
B_{12}^2 P^{-7/3} M_{1.4}^{-2/3} R_6^5 \,\textrm{erg s$^{-1}$} ,
\ee
\noindent
where $R_6$ and $M_{1.4}$ are the neutron star radius in units of $10^6$~cm
and mass in units of 1.4M$_\odot$, respectively, $B_{12}$ is the neutron star
magnetic field strength in units of $10^{12}$~G. Factor $k$ relates the size
of the magnetosphere for a given accretion configuration to the Alfv\'en
radius \citep[$k=0.5$ is usually assumed in the case of the disc accretion;
][]{1979ApJ...234..296G}.

A light curve of \gro\ measured with the {\it Swift}/XRT and {\it Chandra}
telescopes is presented in Fig.\,\ref{fig:lcxrt}. Black circles represent a
bolometric unabsorbed flux measured with the {\it Swift}/XRT telescope under
assumption of the bolometric correction factor of $\simeq2.2$ (see Section
\ref{sec:obs}). From the figure it is clearly seen a dramatic drop (by a
factor around 1000) of the source flux after MJD\,57138.

Note that there is a time gap (about eight days) between the last observation
where the source was significantly detected and the next one where the source
was not detected already. A typical exposure of an individual {\it Swift}/XRT
observation was about only 1 ks. Such a low exposure and large expected
distance to the system didn't allow us to register the source flux. However,
we were able to put upper limits on the source flux in each observation after
the transition of the source to the propeller regime (shown by green
circles in Fig.\ref{fig:lcxrt}). Also we averaged four {\it Swift}/XRT
observations covering time interval MJD\,$57146 - 57162$ with the total
exposure of about 3.5 ks. The source was not detected on the average sky map
with the 2$\sigma$ upper limit comparable with the further {\it Chandra}
detection (see Fig.\,\ref{fig:lcxrt} and recent paper of
\citealt{2018arXiv180910264R} for studying of the low level state). This
drastic change of the source luminosity is very similar to what observed
recently with {\it Swift}/XRT in several other X-ray pulsars
\citep{2016A&A...593A..16T, 2017ApJ...834..209L} and most likely related to
its transition to the propeller regime.  The solid grey line illustrates the
best fit of the flux decay in the tail of the outburst with a Gaussian
function. Although this model is not physically motivated, it fits the
final stages of outbursts before the transition to the propeller regime quite
well \citep[see e.g.,][]{1998ApJ...499L..65C, 2018A&A...610A..46C,
2017ApJ...834..209L}.

\begin{figure*}
\centering
\includegraphics[width=0.9\textwidth,bb=45 260 570 535,clip]{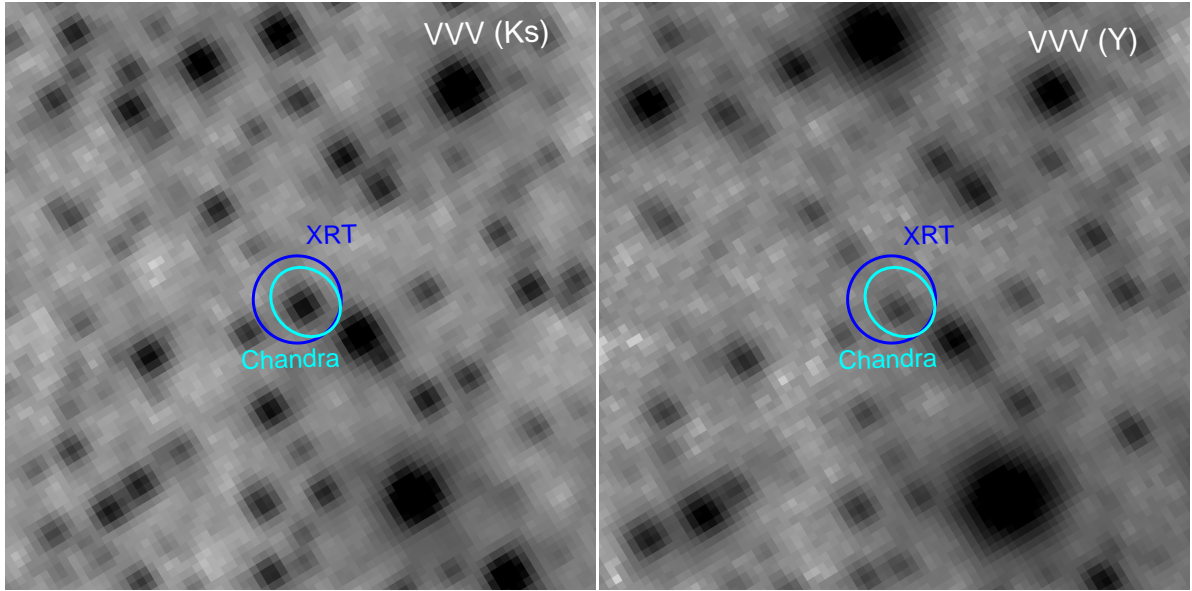}
\caption{Infrared maps of the sky region around \gro\ in $Ks$ and $Y$
filters of the {\it VVV} survey. Localization regions derived from {\it
Swift}/XRT and {\it Chandra} data are shown by the blue circle and cyan
ellipse, respectively.}\label{image}
\end{figure*}

Unfortunately, gaps in the light curve do not allow us to estimate directly
the transition luminosity. However, an abrupt fading of the source flux by
approximately three orders of magnitude on the time scale of several days
after MJD\,57138 is a strong indication that the transition did indeed take
place. A horizontal dashed line in Fig.~\ref{fig:lcxrt} shows an upper limit
$F_{\rm lim}\simeq6\times10^{-11}$ \flux\ on the threshold flux of the
possible transition to the propeller regime.  Substituting $M=1.4 M_{\odot}$,
$R=10$~km and $k=0.5$ to the equation (\ref{eq_prop}) this threshold flux
corresponds to the magnetic field strength $B\simeq(4-5)\times10^{12}$\,G for
distances to the system of 15-20 kpc. Because $F_{\rm lim}$ is just an upper
limit, the magnetic field estimates based on this flux are also should be
considered as upper limits. Note, that these values are in good agreement
with field estimated based on the observed spin-up rate (see
Section\,\ref{sec:pdot}).

\section{X-ray position and spectral properties of \gro}
\label{sec:xray}

An optical companion of \gro\ was not yet known, mainly because of the lack
of an accurate localization of the X-ray source in a very crowded sky field
in the direction to the Galactic Center. Using the {\it Chandra} data we
determined the source position as R.A.= 17$^{\rm h}$49$^{\rm
m}$12.96$^{\rm s}$, \mbox{Dec.=\,-26$^\circ$38\arcmin38.6\arcsec} (J2000) with an
ellipse region uncertainty of $1.6\times1.4$\arcsec\ (90\% confidence level
as provided by {\sc celldetect} routine). We calculated also the enhanced
source position based on the {\it Swift}/XRT data (ObsID.\,0003115012), where
the astrometry is derived using field stars in the UVOT images
\citep{2009MNRAS.397.1177E}\footnote{http://www.swift.ac.uk/user\_objects/index.php}.
The best-fit source position R.A.= 17$^{\rm h}$49$^{\rm m}$12.99$^{\rm s}$,
Dec.=-26$^\circ$38\arcmin38.5\arcsec\ (J2000, error radius 1.9\arcsec, 90\%
confidence level) is in a very good agreement with the {\it Chandra} results.

Thus based on the {\it Swift}/XRT and {\it Chandra} data we have determined,
for the first time, accurate coordinates of \gro\ which will be subsequently
used to search for its infrared counterpart (Fig.\ref{image}).

\begin{figure}
\centering
\includegraphics[width=0.98\columnwidth,bb=40 235 550 700,clip]{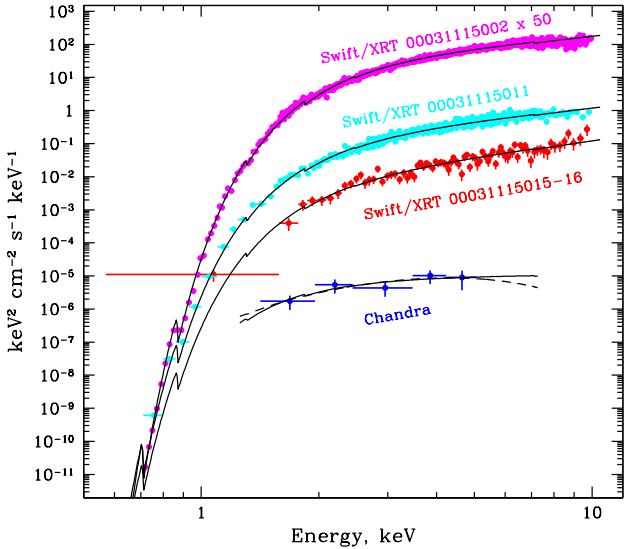}
\caption{Spectra of \gro, obtained with the {\it Swift}/XRT telescope
(magenta, cyan and red circles) and the {\it Chandra} observatory (blue
circles). The spectrum of the first XRT observation is multiplied by a factor
of 50 for clarity. Solid lines represent the best-fit models with the
absorbed powerlaw model. The best-fit model with the black body model for the
low state is shown by dashed line (see text for details).}\label{spec}
\end{figure}

The source spectrum below 10 keV can be well described by a simple power law
model modified by an absorption at low energies. The typical values of the
photon index and absorption are about $\Gamma\simeq0.6-0.8$ and $N_{\rm
H}\simeq(4-5)\times10^{22}$ cm$^{-2}$, respectively, in a wide range of the
source fluxes (we used here both sets of observations in 2008 and 2015,
performed with {\it Swift}/XRT). Several spectra of \gro, obtained in
different states, are shown in Fig.\,\ref{spec}.

As it was shown by \citet{2016MNRAS.463L..46W} and \citet{2016A&A...593A..16T} the
spectra of X-ray pulsars should to become significantly softer after the
transition into the propeller regime and cease of the accretion. The \gro\
spectrum obtained with the {\it Chandra} observatory after such a transition
is shown in Fig.\,\ref{spec} by blue points. It can be described by a black
body model with the temperature of about 1 keV. Formally this spectrum looks
softer in a comparison with the spectra before the transition, but its
temperature is somewhat higher than expected for the propeller regime
($\sim0.5$ keV with the subsequent cooling, \citealt{2016MNRAS.463L..46W}).
It can be associated either with insufficient statistics (formally, the
source spectrum can be also described by the absorbed powerlaw model with the
photon index of $\Gamma\simeq3$), or with a possible continuation of the
accretion after the transition into the propeller regime \citep[see, e.g., a
discussion of different mechanisms of the low state emission
in][]{2017MNRAS.470..126T}.

Finally note, that the absorption value derived from X-ray spectra shows some
excess in a comparison with the value $N_{\rm H} \sim1.1 \times10^{22} $
cm$^{-2}$ given in the standard catalogs LAB \citep {2005A&A...440..775K} and
GASS III \citep{2015A&A...578A..78K}. It can be connected either with an
additional internal absorption due to the matter expelled by a normal star
before and during outbursts or clumps of the interstellar medium on the line
of sight, which are not resolved on the radio maps.

\section{IR counterpart: type and distance estimate}
\label{sec:optic}

\begin{table}
 \centering
  \footnotesize{
  \caption{Visible magnitudes of \gro}\label{table1}
  \begin{tabular}{@{}lc@{}}
  \hline
  $\rm Magnitude_{\rm filter}$ & {\it VVV} values (dr4) \\
 \hline
 $m_Z$    &  $19.65\pm0.17$	    \\
 $m_Y$    &  $17.75\pm0.05$	    \\
 $m_J$    &	 $15.67\pm0.01$		\\
 $m_H$    &	 $14.21\pm0.01$		\\
 $m_{Ks}$ &	 $13.30\pm0.01$ 	\\
\hline
\end{tabular}
}
\end{table}

An accurate localization of \gro\ in X-rays allowed us unambiguously identify
its infrared counterpart based on the {\it VVV}/ESO and {\it UKIDSS}/GPS
surveys. This object is presented in both catalogs with names of
VVV\,J174912.96-263838.92 and UGPS\,J174912.97-263838.9, respectively.
Positions of the object in {\it VVV} and {\it UKIDSS} catalogues are
consistent with each other, so we use the {\it VVV} data further as this
survey covers the sky in a larger number of filters (see
Table\,\ref{table1}). The source coordinates in the {\it VVV} catalogue are
(J2000) R.A.=17$^{\rm h}$49$^{\rm m}$12.968$^{\rm s}$, Dec.=
-26$^\circ$38\arcmin38.93\arcsec\ (Fig.\ref{image}).

An availability of the multiband {\it VVV} photometry gives a possibility to
estimate roughly a spectral energy distribution (SED) of the source and to
compare it with that of other BeXRBs.  Similar approach was used recently
for another distant Be-system \2s\ \citep{2016MNRAS.462.3823L}.

To estimate the SED of the source we need to correct its apparent magnitudes
​​(Table\,\ref{table1}) for the interstellar extinction in each filter. This,
of course, requires a knowledge of an extinction law in the direction to the
source. It is especially important, taking into account preliminary estimates
of the distance to \gro\ as ​​$>12$ kpc
\citep{1997ApJ...488..831S,2009MNRAS.393..419S} and a number of previous
publications, where it was shown that the extinction law for the central
regions of the Galaxy might significantly differ from the standard one
\citep[see, e.g.,][]{2009ApJ...696.1407N,2010A&A...515A..49R,
2010MNRAS.409L..69K, 2015AstL...41..394K,2017ApJ...849L..13A,
2018AstL...44..220K}. For the following analysis we used the absorption law
and extinction ratios in different filters, obtained by
\citet{2017ApJ...849L..13A} from the {\it VVV} survey data for the Galactic
Center direction.

Using these measurements we are able to determine an absorption magnitude and
distance to the Galactic bulge examining the position of red clump giants
(which can be considered as a tracers of the bar -- the cental structure
of the bulge), on the color-magnitude diagram (CMD), reconstructed for all
stars in the vicinity of $\sim2$\arcmin$\times2$\arcmin\ of the studied
object. The observed magnitude and color of the centroid of red clump giants
(RCG) are $m_ {Ks, RCG} = 13.85 \pm 0.05 $ and $(H-Ks)'= 0.84 \pm 0.04 $,
respectively. Knowing an absolute magnitude and unabsorbed color of RCGs in
these filters $M_{Ks, RCG} = -1.61 \pm 0.03$ and $(H-Ks)_0 = 0.11 \pm 0.03$
\citep{2000ApJ...539..732A, 2017AstL...43..545G, 2018AstL...44..220K}, we can
determine the absorption magnitude to RCGs (and subsequently to the Galactic
bulge) as $A_{Ks} = 0.83\pm0.06$ based on the relation $ A_H-A_{Ks} = (H-Ks)
'- (H-Ks)_0 $ and the extinction ratio from \citet{2017ApJ...849L..13A}. It
is known that the distance to RCGs (or in other world to the bar) can
vary at different longitudes \citep[see, e.g.,][]{2012ApJ...744L...8G}. \gro\
is located in $\sim2\deg$ from the Galactic Center, therefore we refined the
distance to the bulge to be more confident in the subsequent calculations.
This distance $d = 8.43\pm0.33$ kpc can be estimated from a relation
$m=M+5log_{10}d -5 + A $, where $m$ is the apparent/measured magnitude, $M$
is the absolute magnitude, $d$ is the distance in pc and $A$ is the
absorption magnitude.

\begin{figure}
\centering
\includegraphics[width=0.98\columnwidth,bb=50 200 580 700,clip]{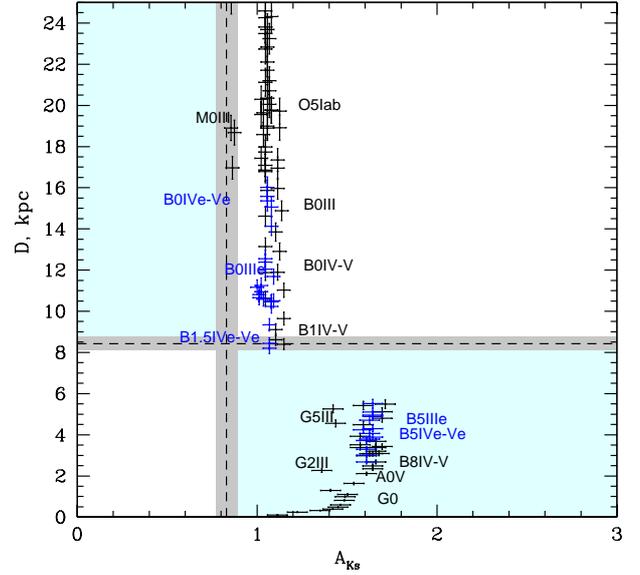}
\caption{'Distance-absorption' diagram showing the distance at which the
counterpart to the source of the corresponding class would be located and
what extinction it would have. The permitted and forbidden classes lie in the
white and blue regions of the diagram, respectively. The dashed lines and
light-gray region indicate the distance and extinction to the bulge and
the accuracy of their estimates.}\label{diag}
\end{figure}

Based on the measurements of the distance and absorption to the Galactic
bulge, we can try to estimate the absorption to the source and its class and
distance. Assuming stars of different spectral and luminosity classes as the
\gro\ companion, we define the correction for the absorption and the distance
required for each of them to satisfy the observed values from the {\it VVV}
survey (Table\,\ref{table1}). It is important to note, that we don't know
exactly where the extinction law become a non-standard one, therefore for
stars which are located before the bulge we use a standard extinction
law $R_{V}=3.1$  \citep{1989ApJ...345..245C},
and for stars in the bulge or behind it the non-standard law is applied
\citep[see][for the detailed description]{2015AstL...41..394K}.

The resulting ``distance-absorption'' diagram is presented in Fig.\ref{diag}.
Stars of different classes, that potentially could be a counterpart of
\gro, lie in the white regions, whereas those that can not be the counterpart
-- in blue ones\footnote{Absolute magnitudes and intrinsic colors of stars of
different spectral and luminosity classes for the corresponding filters were
taken from \citet{2000MNRAS.319..771W, 2006MNRAS.371..185W,
2007MNRAS.374.1549W, 2014AcA....64..261W} and \citet{2015AN....336..159W} for
Be-stars.}. Dashed lines correspond to the absorption and distance to the
Galactic bulge and divide the diagram in these regions.  Two conclusions can
be made from this diagram: 1) a normal star in \gro\ should to have the
class not later then B1-2 and to be located behind the Galactic Center; 2)
the magnitude of the absorption to the object is $A_{Ks} = 1.07 \pm 0.05$.

Knowing the absorption for the filter $Ks$ and extinction laws/ratios, we can
calculate the absorption for other filters and then derive corrected
(unabsorbed) magnitudes for our star. To compare SED for this star with SEDs
for known Be-systems or other stars, the unabsorbed magnitudes were converted
to the absolute ones for two possible distances to the source: 15 and 20
kpc.

\begin{figure}
\centering
\includegraphics[width=0.5\textwidth,bb=50 205 580 700,clip]{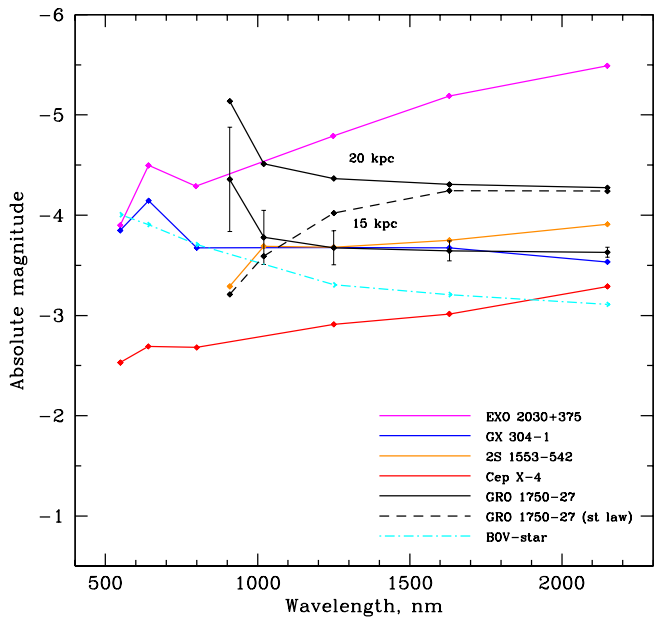}
\caption{Absolute magnitudes of the companion star of \gro\ (black points)
are plotted against wavelength to show its spectral energy distributions for
two possible distance to the source: 15 and 20 kpc. Error bars for magnitudes
are shown only for one curve. SEDs of known BeXRBs EXO\,2030+375, GX\,304-1,
\2s\ and Cep\,X-4 are shown by purple, blue, orange and red points and
curves, respectively and taken from \citet{1997A&AS..126..237C,
2012A&A...539A.114R, 2014MNRAS.445.4235R, 2016MNRAS.462.3823L}. The template
SED for B0V star is shown by the cyan points and dashed-dotted curve for
comparison. Dashed curve demonstrates a possible SED of \gro\ for
the standard absorption law and distance of 15 kpc.}\label{sed}
\end{figure}

The obtained spectral energy distribution of \gro\ (Fig.\ref{sed}) is similar
to SEDs of Be-systems GX\,304-1 and \2s, if \gro\ is located at distances
around 15 kpc (the lower black points and solid curve). At the same time SED
of the B0V star (cyan dashed-dotted curve) is also quite similar to SED of
the \gro\ counterpart for the distance of 12 kpc. Note we do not see in SED
of \gro\ an infrared excess, which is typical for Be-systems due to a
circumstellar disk around the Be star. An absence of such an excess can be
due to IR observations were performed in 2010-2013 when the system was in the
low state. It is known that Be-stars can lose the disk and
spectrometrically look similar to conventional B-stars
\citep{2003A&A...403..239C,2010ApJ...709.1306W,2014ApJ...786..120D}. So we
cannot exclude such a situation for \gro. Note, that a probable uncertainty
in the extinction curve can change SED of the source, but the IR excess is
difficult to reproduce (see dashed line in Fig.\ref{sed} for the standard
absorption law).

Finally, we conclude that the counterpart of \gro\ is a distant star ($>12$
kpc) of an early spectral class, probably a Be-star taking into account its
transient behaviour. This distance estimate is in line with other estimates
discussed above. Spectroscopic observations in the infrared waveband are
required to obtain more robust spectral classification and spectrophotometric
distance to the source.

\section{Conclusions}

In this work the detailed study of the accreting X-ray pulsar \gro\ using
data of X-ray observatories {\it Chandra}, {\it Swift}, and {\it Fermi}, and
the near-infrared survey {\it VVV}/ESO is presented.

We have measured, for the first time, accurate X-ray coordinates of \gro\ (J2000)
R.A.= 17$^{\rm h}$49$^{\rm m}$12.99$^{\rm s}$,
Dec.=-26$^\circ$38\arcmin38.5\arcsec\ and determined its infrared
counterpart.

A comparison of the spectral energy distribution of \gro\ with those of known
Be/X-ray binaries and early type stars has allowed us to estimate a lower
distance limit to the source at $>12$ kpc. Moreover, the analysis of the
observed spin-up of the pulsar during the giant outburst in 2015 provides an
independent distance estimate of $14-22$\,kpc, and also allows to estimate the
magnetic field on the surface of the neutron star as
\mbox{$B\simeq(3.5-4.5)\times10^{12}$\,G.}

Finally, the monitoring of the source with the {\it Swift}/XRT telescope
revealed a rapid drop of the source flux in the end of the 2015 outburst,
likely associated with the transition to the propeller regime. A detection of
this transition gives an estimate of the magnetic field of the neutron star
as $B\simeq(4-5)\times10^{12}$\,G for distances to the system of 15-20 kpc, that
is well agreed with above results from the infrared and spin-up analysis.

\section*{Acknowledgments}

This work was supported by the Russian Foundation of Basic Research (grant
17-52-80139 BRICS-a). SST acknowledges the support of the Academy of Finland
grants 309228, 316932 and 317552. V.D. thank the Deutsches Zentrum for Luft-
und Raumfahrt (DLR) and Deutsche Forschungsgemeinschaft (DFG) for the
support. Authors thank to Prof. Frederick Walter for comments and suggestions
that allowed to improve the manuscript. The research has made use of {\it
Chandra} data and software provided by the Chandra X-ray Center. We used also
data supplied by the UK Swift Science Data Centre at the University of
Leicester. Results are also based on data products from \textit{VVV} Survey
observations made with the {\it VISTA} telescope at the ESO Paranal
Observatory.

\bibliographystyle{mnras}
\bibliography{groj1750_bib}

\begin{thebibliography}{}
\makeatletter
\relax
\def\mn@urlcharsother{\let\do\@makeother \do\$\do\&\do\#\do\^\do\_\do\%\do\~}
\def\mn@doi{\begingroup\mn@urlcharsother \@ifnextchar [ {\mn@doi@}
  {\mn@doi@[]}}
\def\mn@doi@[#1]#2{\def\@tempa{#1}\ifx\@tempa\@empty \href
  {http://dx.doi.org/#2} {doi:#2}\else \href {http://dx.doi.org/#2} {#1}\fi
  \endgroup}
\def\mn@eprint#1#2{\mn@eprint@#1:#2::\@nil}
\def\mn@eprint@arXiv#1{\href {http://arxiv.org/abs/#1} {{\tt arXiv:#1}}}
\def\mn@eprint@dblp#1{\href {http://dblp.uni-trier.de/rec/bibtex/#1.xml}
  {dblp:#1}}
\def\mn@eprint@#1:#2:#3:#4\@nil{\def\@tempa {#1}\def\@tempb {#2}\def\@tempc
  {#3}\ifx \@tempc \@empty \let \@tempc \@tempb \let \@tempb \@tempa \fi \ifx
  \@tempb \@empty \def\@tempb {arXiv}\fi \@ifundefined
  {mn@eprint@\@tempb}{\@tempb:\@tempc}{\expandafter \expandafter \csname
  mn@eprint@\@tempb\endcsname \expandafter{\@tempc}}}

\bibitem[\protect\citeauthoryear{{Alonso-Garc{\'{\i}}a}
  et~al.,}{{Alonso-Garc{\'{\i}}a} et~al.}{2017}]{2017ApJ...849L..13A}
{Alonso-Garc{\'{\i}}a} J.,  et~al., 2017, \mn@doi [\apjl]
  {10.3847/2041-8213/aa92c3}, \href
  {http://adsabs.harvard.edu/abs/2017ApJ...849L..13A} {849, L13}

\bibitem[\protect\citeauthoryear{{Alves}}{{Alves}}{2000}]{2000ApJ...539..732A}
{Alves} D.~R.,  2000, \mn@doi [\apj] {10.1086/309278}, \href
  {http://adsabs.harvard.edu/abs/2000ApJ...539..732A} {539, 732}

\bibitem[\protect\citeauthoryear{{Campana}, {Stella}, {Mereghetti}, {Colpi},
  {Tavani}, {Ricci}, {Dal Fiume}  \& {Belloni}}{{Campana}
  et~al.}{1998}]{1998ApJ...499L..65C}
{Campana} S.,  {Stella} L.,  {Mereghetti} S.,  {Colpi} M.,  {Tavani} M.,
  {Ricci} D.,  {Dal Fiume} D.,   {Belloni} T.,  1998, \mn@doi [\apjl]
  {10.1086/311357}, \href {http://adsabs.harvard.edu/abs/1998ApJ...499L..65C}
  {499, L65}

\bibitem[\protect\citeauthoryear{{Campana}, {Gastaldello}, {Stella}, {Israel},
  {Colpi}, {Pizzolato}, {Orlandini}  \& {Dal Fiume}}{{Campana}
  et~al.}{2001}]{2001ApJ...561..924C}
{Campana} S.,  {Gastaldello} F.,  {Stella} L.,  {Israel} G.~L.,  {Colpi} M.,
  {Pizzolato} F.,  {Orlandini} M.,   {Dal Fiume} D.,  2001, \mn@doi [\apj]
  {10.1086/323317}, \href {http://adsabs.harvard.edu/abs/2001ApJ...561..924C}
  {561, 924}

\bibitem[\protect\citeauthoryear{{Campana}, {Stella}  \& {Kennea}}{{Campana}
  et~al.}{2008}]{2008ApJ...684L..99C}
{Campana} S.,  {Stella} L.,   {Kennea} J.~A.,  2008, \mn@doi [\apjl]
  {10.1086/592002}, \href {http://adsabs.harvard.edu/abs/2008ApJ...684L..99C}
  {684, L99}

\bibitem[\protect\citeauthoryear{{Campana}, {Stella}, {Mereghetti}  \& {de
  Martino}}{{Campana} et~al.}{2018}]{2018A&A...610A..46C}
{Campana} S.,  {Stella} L.,  {Mereghetti} S.,   {de Martino} D.,  2018, \mn@doi
  [\aap] {10.1051/0004-6361/201730769}, \href
  {http://adsabs.harvard.edu/abs/2018A%26A...610A..46C} {610, A46}

\bibitem[\protect\citeauthoryear{{Cardelli}, {Clayton}  \& {Mathis}}{{Cardelli}
  et~al.}{1989}]{1989ApJ...345..245C}
{Cardelli} J.~A.,  {Clayton} G.~C.,   {Mathis} J.~S.,  1989, \mn@doi [\apj]
  {10.1086/167900}, \href {http://adsabs.harvard.edu/abs/1989ApJ...345..245C}
  {345, 245}

\bibitem[\protect\citeauthoryear{{Clark}, {Tarasov}  \& {Panko}}{{Clark}
  et~al.}{2003}]{2003A&A...403..239C}
{Clark} J.~S.,  {Tarasov} A.~E.,   {Panko} E.~A.,  2003, \mn@doi [\aap]
  {10.1051/0004-6361:20030248}, \href
  {http://adsabs.harvard.edu/abs/2003A%26A...403..239C} {403, 239}

\bibitem[\protect\citeauthoryear{{Coe}, {Buckley}, {Fabregat}, {Steele},
  {Still}  \& {Torrejon}}{{Coe} et~al.}{1997}]{1997A&AS..126..237C}
{Coe} M.~J.,  {Buckley} D.~A.~H.,  {Fabregat} J.,  {Steele} L.~A.,  {Still}
  M.~D.,   {Torrejon} J.~M.,  1997, \mn@doi [\aaps] {10.1051/aas:1997260},
  \href {http://adsabs.harvard.edu/abs/1997A%26AS..126..237C} {126, 237}

\bibitem[\protect\citeauthoryear{{Cui}}{{Cui}}{1997}]{1997ApJ...482L.163C}
{Cui} W.,  1997, \mn@doi [\apjl] {10.1086/310712}, \href
  {http://adsabs.harvard.edu/abs/1997ApJ...482L.163C} {482, L163}

\bibitem[\protect\citeauthoryear{{Dotani}, {Fujimoto}, {Nagase}  \&
  {Inoue}}{{Dotani} et~al.}{1995}]{1995IAUC.6241....2D}
{Dotani} T.,  {Fujimoto} R.,  {Nagase} F.,   {Inoue} H.,  1995, \iaucirc, \href
  {http://adsabs.harvard.edu/abs/1995IAUC.6241....2D} {6241}

\bibitem[\protect\citeauthoryear{{Draper}, {Wisniewski}, {Bjorkman}, {Meade},
  {Haubois}, {Mota}, {Carciofi}  \& {Bjorkman}}{{Draper}
  et~al.}{2014}]{2014ApJ...786..120D}
{Draper} Z.~H.,  {Wisniewski} J.~P.,  {Bjorkman} K.~S.,  {Meade} M.~R.,
  {Haubois} X.,  {Mota} B.~C.,  {Carciofi} A.~C.,   {Bjorkman} J.~E.,  2014,
  \mn@doi [\apj] {10.1088/0004-637X/786/2/120}, \href
  {http://adsabs.harvard.edu/abs/2014ApJ...786..120D} {786, 120}

\bibitem[\protect\citeauthoryear{{Evans} et~al.,}{{Evans}
  et~al.}{2009}]{2009MNRAS.397.1177E}
{Evans} P.~A.,  et~al., 2009, \mn@doi [\mnras]
  {10.1111/j.1365-2966.2009.14913.x}, \href
  {http://adsabs.harvard.edu/abs/2009MNRAS.397.1177E} {397, 1177}

\bibitem[\protect\citeauthoryear{{Filippova}, {Mereminskiy}, {Lutovinov},
  {Molkov}  \& {Tsygankov}}{{Filippova} et~al.}{2017}]{2017AstL...43..706F}
{Filippova} E.~V.,  {Mereminskiy} I.~A.,  {Lutovinov} A.~A.,  {Molkov} S.~V.,
  {Tsygankov} S.~S.,  2017, \mn@doi [Astronomy Letters]
  {10.1134/S1063773717110020}, \href
  {http://adsabs.harvard.edu/abs/2017AstL...43..706F} {43, 706}

\bibitem[\protect\citeauthoryear{{Gerhard} \& {Martinez-Valpuesta}}{{Gerhard}
  \& {Martinez-Valpuesta}}{2012}]{2012ApJ...744L...8G}
{Gerhard} O.,  {Martinez-Valpuesta} I.,  2012, \mn@doi [\apjl]
  {10.1088/2041-8205/744/1/L8}, \href
  {http://adsabs.harvard.edu/abs/2012ApJ...744L...8G} {744, L8}

\bibitem[\protect\citeauthoryear{{Ghosh} \& {Lamb}}{{Ghosh} \&
  {Lamb}}{1979}]{1979ApJ...234..296G}
{Ghosh} P.,  {Lamb} F.~K.,  1979, \mn@doi [\apj] {10.1086/157498}, \href
  {http://adsabs.harvard.edu/abs/1979ApJ...234..296G} {234, 296}

\bibitem[\protect\citeauthoryear{{Gontcharov}}{{Gontcharov}}{2017}]{2017AstL...43..545G}
{Gontcharov} G.~A.,  2017, \mn@doi [Astronomy Letters]
  {10.1134/S1063773717060044}, \href
  {http://adsabs.harvard.edu/abs/2017AstL...43..545G} {43, 545}

\bibitem[\protect\citeauthoryear{{Illarionov} \& {Sunyaev}}{{Illarionov} \&
  {Sunyaev}}{1975}]{1975A&A....39..185I}
{Illarionov} A.~F.,  {Sunyaev} R.~A.,  1975, \aap, \href
  {http://adsabs.harvard.edu/abs/1975A%26A....39..185I} {39, 185}

\bibitem[\protect\citeauthoryear{{Kalberla} \& {Haud}}{{Kalberla} \&
  {Haud}}{2015}]{2015A&A...578A..78K}
{Kalberla} P.~M.~W.,  {Haud} U.,  2015, \mn@doi [\aap]
  {10.1051/0004-6361/201525859}, \href
  {http://adsabs.harvard.edu/abs/2015A%26A...578A..78K} {578, A78}

\bibitem[\protect\citeauthoryear{{Kalberla}, {Burton}, {Hartmann}, {Arnal},
  {Bajaja}, {Morras}  \& {P{\"o}ppel}}{{Kalberla}
  et~al.}{2005}]{2005A&A...440..775K}
{Kalberla} P.~M.~W.,  {Burton} W.~B.,  {Hartmann} D.,  {Arnal} E.~M.,  {Bajaja}
  E.,  {Morras} R.,   {P{\"o}ppel} W.~G.~L.,  2005, \mn@doi [\aap]
  {10.1051/0004-6361:20041864}, \href
  {http://adsabs.harvard.edu/abs/2005A%26A...440..775K} {440, 775}

\bibitem[\protect\citeauthoryear{{Karasev} \& {Lutovinov}}{{Karasev} \&
  {Lutovinov}}{2018}]{2018AstL...44..220K}
{Karasev} D.~I.,  {Lutovinov} A.~A.,  2018, \mn@doi [Astronomy Letters]
  {10.1134/S1063773718040047}, \href
  {http://adsabs.harvard.edu/abs/2018AstL...44..220K} {44, 220}

\bibitem[\protect\citeauthoryear{{Karasev}, {Lutovinov}  \&
  {Burenin}}{{Karasev} et~al.}{2010}]{2010MNRAS.409L..69K}
{Karasev} D.~I.,  {Lutovinov} A.~A.,   {Burenin} R.~A.,  2010, \mn@doi [\mnras]
  {10.1111/j.1745-3933.2010.00949.x}, \href
  {http://adsabs.harvard.edu/abs/2010MNRAS.409L..69K} {409, L69}

\bibitem[\protect\citeauthoryear{{Karasev}, {Tsygankov}  \&
  {Lutovinov}}{{Karasev} et~al.}{2015}]{2015AstL...41..394K}
{Karasev} D.~I.,  {Tsygankov} S.~S.,   {Lutovinov} A.~A.,  2015, \mn@doi
  [Astronomy Letters] {10.1134/S1063773715080022}, \href
  {http://adsabs.harvard.edu/abs/2015AstL...41..394K} {41, 394}

\bibitem[\protect\citeauthoryear{{Klu{\'z}niak} \& {Rappaport}}{{Klu{\'z}niak}
  \& {Rappaport}}{2007}]{2007ApJ...671.1990K}
{Klu{\'z}niak} W.,  {Rappaport} S.,  2007, \mn@doi [\apj] {10.1086/522954},
  \href {http://adsabs.harvard.edu/abs/2007ApJ...671.1990K} {671, 1990}

\bibitem[\protect\citeauthoryear{{Koh}, {Chakrabarty}, {Prince}, {Vaughan},
  {Zhang}, {Scott}, {Finger}  \& {Wilson}}{{Koh}
  et~al.}{1995}]{1995IAUC.6222....1K}
{Koh} T.,  {Chakrabarty} D.,  {Prince} T.~A.,  {Vaughan} B.,  {Zhang} S.~N.,
  {Scott} M.,  {Finger} M.~H.,   {Wilson} R.~B.,  1995, \iaucirc, \href
  {http://adsabs.harvard.edu/abs/1995IAUC.6222....1K} {6222}

\bibitem[\protect\citeauthoryear{{Krimm} et~al.,}{{Krimm}
  et~al.}{2013}]{2013ApJS..209...14K}
{Krimm} H.~A.,  et~al., 2013, \mn@doi [\apjs] {10.1088/0067-0049/209/1/14},
  \href {http://adsabs.harvard.edu/abs/2013ApJS..209...14K} {209, 14}

\bibitem[\protect\citeauthoryear{{Lutovinov}, {Buckley}, {Townsend},
  {Tsygankov}  \& {Kennea}}{{Lutovinov} et~al.}{2016}]{2016MNRAS.462.3823L}
{Lutovinov} A.~A.,  {Buckley} D.~A.~H.,  {Townsend} L.~J.,  {Tsygankov} S.~S.,
   {Kennea} J.,  2016, \mn@doi [\mnras] {10.1093/mnras/stw1889}, \href
  {http://adsabs.harvard.edu/abs/2016MNRAS.462.3823L} {462, 3823}

\bibitem[\protect\citeauthoryear{{Lutovinov}, {Tsygankov}, {Krivonos}, {Molkov}
   \& {Poutanen}}{{Lutovinov} et~al.}{2017}]{2017ApJ...834..209L}
{Lutovinov} A.~A.,  {Tsygankov} S.~S.,  {Krivonos} R.~A.,  {Molkov} S.~V.,
  {Poutanen} J.,  2017, \mn@doi [\apj] {10.3847/1538-4357/834/2/209}, \href
  {http://adsabs.harvard.edu/abs/2017ApJ...834..209L} {834, 209}

\bibitem[\protect\citeauthoryear{{Meegan} et~al.,}{{Meegan}
  et~al.}{2009}]{2009ApJ...702..791M}
{Meegan} C.,  et~al., 2009, \mn@doi [\apj] {10.1088/0004-637X/702/1/791}, \href
  {http://adsabs.harvard.edu/abs/2009ApJ...702..791M} {702, 791}

\bibitem[\protect\citeauthoryear{{Negueruela} \& {Okazaki}}{{Negueruela} \&
  {Okazaki}}{2001}]{2001A&A...369..108N}
{Negueruela} I.,  {Okazaki} A.~T.,  2001, \mn@doi [\aap]
  {10.1051/0004-6361:20010146}, \href
  {http://adsabs.harvard.edu/abs/2001A%26A...369..108N} {369, 108}

\bibitem[\protect\citeauthoryear{{Nishiyama}, {Tamura}, {Hatano}, {Kato},
  {Tanab{\'e}}, {Sugitani}  \& {Nagata}}{{Nishiyama}
  et~al.}{2009}]{2009ApJ...696.1407N}
{Nishiyama} S.,  {Tamura} M.,  {Hatano} H.,  {Kato} D.,  {Tanab{\'e}} T.,
  {Sugitani} K.,   {Nagata} T.,  2009, \mn@doi [\apj]
  {10.1088/0004-637X/696/2/1407}, \href
  {http://adsabs.harvard.edu/abs/2009ApJ...696.1407N} {696, 1407}

\bibitem[\protect\citeauthoryear{{Okazaki} \& {Negueruela}}{{Okazaki} \&
  {Negueruela}}{2001}]{2001A&A...377..161O}
{Okazaki} A.~T.,  {Negueruela} I.,  2001, \mn@doi [\aap]
  {10.1051/0004-6361:20011083}, \href
  {http://adsabs.harvard.edu/abs/2001A%26A...377..161O} {377, 161}

\bibitem[\protect\citeauthoryear{{Okazaki}, {Hayasaki}  \&
  {Moritani}}{{Okazaki} et~al.}{2013}]{2013PASJ...65...41O}
{Okazaki} A.~T.,  {Hayasaki} K.,   {Moritani} Y.,  2013, \mn@doi [\pasj]
  {10.1093/pasj/65.2.41}, \href
  {http://adsabs.harvard.edu/abs/2013PASJ...65...41O} {65, 41}

\bibitem[\protect\citeauthoryear{{Parfrey}, {Spitkovsky}  \&
  {Beloborodov}}{{Parfrey} et~al.}{2016}]{2016ApJ...822...33P}
{Parfrey} K.,  {Spitkovsky} A.,   {Beloborodov} A.~M.,  2016, \mn@doi [\apj]
  {10.3847/0004-637X/822/1/33}, \href
  {http://adsabs.harvard.edu/abs/2016ApJ...822...33P} {822, 33}

\bibitem[\protect\citeauthoryear{{Poutanen}, {Mushtukov}, {Suleimanov},
  {Tsygankov}, {Nagirner}, {Doroshenko}  \& {Lutovinov}}{{Poutanen}
  et~al.}{2013}]{2013ApJ...777..115P}
{Poutanen} J.,  {Mushtukov} A.~A.,  {Suleimanov} V.~F.,  {Tsygankov} S.~S.,
  {Nagirner} D.~I.,  {Doroshenko} V.,   {Lutovinov} A.~A.,  2013, \mn@doi
  [\apj] {10.1088/0004-637X/777/2/115}, \href
  {http://adsabs.harvard.edu/abs/2013ApJ...777..115P} {777, 115}

\bibitem[\protect\citeauthoryear{{Reig}}{{Reig}}{2011}]{2011Ap&SS.332....1R}
{Reig} P.,  2011, \mn@doi [\apss] {10.1007/s10509-010-0575-8}, \href
  {http://adsabs.harvard.edu/abs/2011Ap%26SS.332....1R} {332, 1}

\bibitem[\protect\citeauthoryear{{Reig}, {Blinov}, {Papadakis}, {Kylafis}  \&
  {Tassis}}{{Reig} et~al.}{2014}]{2014MNRAS.445.4235R}
{Reig} P.,  {Blinov} D.,  {Papadakis} I.,  {Kylafis} N.,   {Tassis} K.,  2014,
  \mn@doi [\mnras] {10.1093/mnras/stu2322}, \href
  {http://adsabs.harvard.edu/abs/2014MNRAS.445.4235R} {445, 4235}

\bibitem[\protect\citeauthoryear{{Revnivtsev}, {van den Berg}, {Burenin},
  {Grindlay}, {Karasev}  \& {Forman}}{{Revnivtsev}
  et~al.}{2010}]{2010A&A...515A..49R}
{Revnivtsev} M.,  {van den Berg} M.,  {Burenin} R.,  {Grindlay} J.~E.,
  {Karasev} D.,   {Forman} W.,  2010, \mn@doi [\aap]
  {10.1051/0004-6361/200913527}, \href
  {http://adsabs.harvard.edu/abs/2010A%26A...515A..49R} {515, A49}

\bibitem[\protect\citeauthoryear{{Riquelme}, {Torrej{\'o}n}  \&
  {Negueruela}}{{Riquelme} et~al.}{2012}]{2012A&A...539A.114R}
{Riquelme} M.~S.,  {Torrej{\'o}n} J.~M.,   {Negueruela} I.,  2012, \mn@doi
  [\aap] {10.1051/0004-6361/201117738}, \href
  {http://adsabs.harvard.edu/abs/2012A%26A...539A.114R} {539, A114}

\bibitem[\protect\citeauthoryear{{Rouco Escorial}, {Wijnands}, {Ootes},
  {Degenaar}, {Snelders}, {Kaper}, {Cackett}  \& {Homan}}{{Rouco Escorial}
  et~al.}{2018}]{2018arXiv180910264R}
{Rouco Escorial} A.,  {Wijnands} R.,  {Ootes} L.~S.,  {Degenaar} N.,
  {Snelders} M.,  {Kaper} L.,  {Cackett} E.~M.,   {Homan} J.,  2018, arXiv
  e-prints, \href {http://adsabs.harvard.edu/abs/2018arXiv180910264R} {1809.10264}

\bibitem[\protect\citeauthoryear{{Scott}, {Finger}, {Wilson}, {Koh}, {Prince},
  {Vaughan}  \& {Chakrabarty}}{{Scott} et~al.}{1997}]{1997ApJ...488..831S}
{Scott} D.~M.,  {Finger} M.~H.,  {Wilson} R.~B.,  {Koh} D.~T.,  {Prince} T.~A.,
   {Vaughan} B.~A.,   {Chakrabarty} D.,  1997, \mn@doi [\apj] {10.1086/304740},
  \href {http://adsabs.harvard.edu/abs/1997ApJ...488..831S} {488, 831}

\bibitem[\protect\citeauthoryear{{Shaw}, {Hill}, {Kuulkers}, {Brandt},
  {Chenevez}  \& {Kretschmar}}{{Shaw} et~al.}{2009}]{2009MNRAS.393..419S}
{Shaw} S.~E.,  {Hill} A.~B.,  {Kuulkers} E.,  {Brandt} S.,  {Chenevez} J.,
  {Kretschmar} P.,  2009, \mn@doi [\mnras] {10.1111/j.1365-2966.2008.14212.x},
  \href {http://adsabs.harvard.edu/abs/2009MNRAS.393..419S} {393, 419}

\bibitem[\protect\citeauthoryear{{Stella}, {White}  \& {Rosner}}{{Stella}
  et~al.}{1986}]{1986ApJ...308..669S}
{Stella} L.,  {White} N.~E.,   {Rosner} R.,  1986, \mn@doi [\apj]
  {10.1086/164538}, \href {http://adsabs.harvard.edu/abs/1986ApJ...308..669S}
  {308, 669}

\bibitem[\protect\citeauthoryear{{Tsygankov}, {Mushtukov}, {Suleimanov}  \&
  {Poutanen}}{{Tsygankov} et~al.}{2016a}]{2016MNRAS.457.1101T}
{Tsygankov} S.~S.,  {Mushtukov} A.~A.,  {Suleimanov} V.~F.,   {Poutanen} J.,
  2016a, \mn@doi [\mnras] {10.1093/mnras/stw046}, \href
  {http://adsabs.harvard.edu/abs/2016MNRAS.457.1101T} {457, 1101}

\bibitem[\protect\citeauthoryear{{Tsygankov}, {Lutovinov}, {Doroshenko},
  {Mushtukov}, {Suleimanov}  \& {Poutanen}}{{Tsygankov}
  et~al.}{2016b}]{2016A&A...593A..16T}
{Tsygankov} S.~S.,  {Lutovinov} A.~A.,  {Doroshenko} V.,  {Mushtukov} A.~A.,
  {Suleimanov} V.,   {Poutanen} J.,  2016b, \mn@doi [\aap]
  {10.1051/0004-6361/201628236}, \href
  {http://adsabs.harvard.edu/abs/2016A%26A...593A..16T} {593, A16}

\bibitem[\protect\citeauthoryear{{Tsygankov}, {Wijnands}, {Lutovinov},
  {Degenaar}  \& {Poutanen}}{{Tsygankov} et~al.}{2017a}]{2017MNRAS.470..126T}
{Tsygankov} S.~S.,  {Wijnands} R.,  {Lutovinov} A.~A.,  {Degenaar} N.,
  {Poutanen} J.,  2017a, \mn@doi [\mnras] {10.1093/mnras/stx1255}, \href
  {http://adsabs.harvard.edu/abs/2017MNRAS.470..126T} {470, 126}

\bibitem[\protect\citeauthoryear{{Tsygankov}, {Doroshenko}, {Lutovinov},
  {Mushtukov}  \& {Poutanen}}{{Tsygankov} et~al.}{2017b}]{2017A&A...605A..39T}
{Tsygankov} S.~S.,  {Doroshenko} V.,  {Lutovinov} A.~A.,  {Mushtukov} A.~A.,
  {Poutanen} J.,  2017b, \mn@doi [\aap] {10.1051/0004-6361/201730553}, \href
  {http://adsabs.harvard.edu/abs/2017A%26A...605A..39T} {605, A39}

\bibitem[\protect\citeauthoryear{{Tsygankov}, {Mushtukov}, {Suleimanov},
  {Doroshenko}, {Abolmasov}, {Lutovinov}  \& {Poutanen}}{{Tsygankov}
  et~al.}{2017c}]{2017A&A...608A..17T}
{Tsygankov} S.~S.,  {Mushtukov} A.~A.,  {Suleimanov} V.~F.,  {Doroshenko} V.,
  {Abolmasov} P.~K.,  {Lutovinov} A.~A.,   {Poutanen} J.,  2017c, \mn@doi
  [\aap] {10.1051/0004-6361/201630248}, \href
  {http://adsabs.harvard.edu/abs/2017A%26A...608A..17T} {608, A17}

\bibitem[\protect\citeauthoryear{{Wachter}, {Leach}  \& {Kellogg}}{{Wachter}
  et~al.}{1979}]{1979ApJ...230..274W}
{Wachter} K.,  {Leach} R.,   {Kellogg} E.,  1979, \mn@doi [\apj]
  {10.1086/157084}, \href {http://adsabs.harvard.edu/abs/1979ApJ...230..274W}
  {230, 274}

\bibitem[\protect\citeauthoryear{{Walter}, {Lutovinov}, {Bozzo}  \&
  {Tsygankov}}{{Walter} et~al.}{2015}]{2015A&ARv..23....2W}
{Walter} R.,  {Lutovinov} A.~A.,  {Bozzo} E.,   {Tsygankov} S.~S.,  2015,
  \mn@doi [\aapr] {10.1007/s00159-015-0082-6}, \href
  {http://adsabs.harvard.edu/abs/2015A%26ARv..23....2W} {23, 2}

\bibitem[\protect\citeauthoryear{{Wang}}{{Wang}}{1995}]{1995ApJ...449L.153W}
{Wang} Y.-M.,  1995, \mn@doi [\apjl] {10.1086/309649}, \href
  {http://adsabs.harvard.edu/abs/1995ApJ...449L.153W} {449, L153}

\bibitem[\protect\citeauthoryear{{Wegner}}{{Wegner}}{2000}]{2000MNRAS.319..771W}
{Wegner} W.,  2000, \mn@doi [\mnras] {10.1046/j.1365-8711.2000.03884.x}, \href
  {http://adsabs.harvard.edu/abs/2000MNRAS.319..771W} {319, 771}

\bibitem[\protect\citeauthoryear{{Wegner}}{{Wegner}}{2006}]{2006MNRAS.371..185W}
{Wegner} W.,  2006, \mn@doi [\mnras] {10.1111/j.1365-2966.2006.10549.x}, \href
  {http://adsabs.harvard.edu/abs/2006MNRAS.371..185W} {371, 185}

\bibitem[\protect\citeauthoryear{{Wegner}}{{Wegner}}{2007}]{2007MNRAS.374.1549W}
{Wegner} W.,  2007, \mn@doi [\mnras] {10.1111/j.1365-2966.2006.11265.x}, \href
  {http://adsabs.harvard.edu/abs/2007MNRAS.374.1549W} {374, 1549}

\bibitem[\protect\citeauthoryear{{Wegner}}{{Wegner}}{2014}]{2014AcA....64..261W}
{Wegner} W.,  2014, \actaa, \href
  {http://adsabs.harvard.edu/abs/2014AcA....64..261W} {64, 261}

\bibitem[\protect\citeauthoryear{{Wegner}}{{Wegner}}{2015}]{2015AN....336..159W}
{Wegner} W.,  2015, \mn@doi [Astronomische Nachrichten]
  {10.1002/asna.201312143}, \href
  {http://adsabs.harvard.edu/abs/2015AN....336..159W} {336, 159}

\bibitem[\protect\citeauthoryear{{Wijnands} \& {Degenaar}}{{Wijnands} \&
  {Degenaar}}{2016}]{2016MNRAS.463L..46W}
{Wijnands} R.,  {Degenaar} N.,  2016, \mn@doi [\mnras] {10.1093/mnrasl/slw096},
  \href {http://adsabs.harvard.edu/abs/2016MNRAS.463L..46W} {463, L46}

\bibitem[\protect\citeauthoryear{{Wisniewski}, {Draper}, {Bjorkman}, {Meade},
  {Bjorkman}  \& {Kowalski}}{{Wisniewski} et~al.}{2010}]{2010ApJ...709.1306W}
{Wisniewski} J.~P.,  {Draper} Z.~H.,  {Bjorkman} K.~S.,  {Meade} M.~R.,
  {Bjorkman} J.~E.,   {Kowalski} A.~F.,  2010, \mn@doi [\apj]
  {10.1088/0004-637X/709/2/1306}, \href
  {http://adsabs.harvard.edu/abs/2010ApJ...709.1306W} {709, 1306}

\makeatother
\end{thebibliography}

\label{lastpage}
\end{document}